\documentclass[10pt,conference]{IEEEtran}
\IEEEoverridecommandlockouts

\usepackage{cite}
\usepackage[utf8]{inputenc}
\usepackage{amsmath,amssymb,amsfonts}
\usepackage{algorithmic}
\usepackage{algorithm}
\usepackage{graphicx}
\usepackage{textcomp}
\usepackage{xcolor}
\usepackage{balance}
\usepackage{makecell}
\usepackage{pifont}
\usepackage{url}
\usepackage{flushend}

\usepackage{hyperref}
\DeclareRobustCommand{\myurl}[1]{\url{#1}}

\def\BibTeX{{\rm B\kern-.05em{\sc i\kern-.025em b}\kern-.08em T\kern-.1667em\lower.7ex\hbox{E}\kern-.125emX}}

\newcommand{\x}{\mathbf{x}}
\newcommand{\y}{\mathbf{y}}

\newcommand{\X}{\mathbf{X}}
\newcommand{\Y}{\mathbf{Y}}

\newcommand{\A}{\mathbf{A}}

\newcommand{\G}{\mathbf{G}}
\newcommand{\task}{\mathcal{T}}
\newcommand{\gvec}{\mathbf{g}}
\newcommand{\avec}{\mathbf{a}}

\newcommand{\frob}[1]{\|{#1}\|_\mathrm{F}}
\newcommand{\eye}[1]{\mathbf{I}_{#1}}

\def \SemL{{\mathcal{L}_\mathrm{S}}}
\def \EmL{{\mathcal{L}_\mathrm{E}}}

\def \Tdim{{\theta}}
\def \Tdimc{{\overline{\theta}}}
\def \Tlatc{\X}
\def \Tlatr{\mathbf{s}_{\mathrm{T}}}

\def \Rdim{{\omega}}
\def \Rdimc{{\overline{\omega}}}
\def \Rlatc{\Y}
\def \Rlatr{\mathbf{s}_{\mathrm{R}}}

\def \reg{{\gamma}}
\def \AlignMat{{\mathbf \A}}
\def \AlignMatL{{\AlignMat_\mathrm{L}}}
\def \AlignMatF{{\AlignMat_\mathrm{F}}}

\def \simmodule{{g_{\boldsymbol{\xi}}}}

\def \H{\mathbf{H}}

\def \equalizer{\mathbf{Q}}
\def\Nt{{{N}_{\mathrm{T}}}}
\def\Nr{{{N}_{\mathrm{R}}}}
\def\snr{{\text{SNR}}}
\def\snrdb{{\text{SNR}_\text{[dB]}}}

\def \nclusters{\kappa}
\def \nsamples{\varrho}
\def \cluster{\mathcal{C}}

\newcommand{\W}{\mathbf{W}}
\newcommand{\Ups}{\mathbf{\Upsilon}}
\def \amplconst{{\phi}}

\begin{document}

\title{Over-the-Air Semantic Alignment with\\ Stacked Intelligent Metasurfaces 

\thanks{
This work has been supported by the Smart Networks and Services
Joint Undertaking projects 6G-DISAC and 6G-GOALS under
the European Union's Horizon Europe research and innovation programme
under Grant Agreement numbers 101139130, and 101139232 respectively.}}

\author{Mario Edoardo Pandolfo$^{1,2}$, Kyriakos Stylianopoulos$^3$, George C. Alexandropoulos$^3$, and Paolo Di Lorenzo$^{2,4}$ \smallskip\\
$^1$DIAG Department, Sapienza University of Rome, via Ariosto 25, Rome, Italy\\
$^2$National Inter-University Consortium for Telecommunications (CNIT), Parma, Italy\\
$^3$Informatics and Telecommunications Department, National and Kapodistrian University of Athens, Greece\\
$^4$DIET Department, Sapienza University of Rome, Via Eudossiana 18, Rome, Italy 
\\
e-mails:  \{marioedoardo.pandolfo,paolo.dilorenzo\}@uniroma1.it, \{kstylianop,alexandg\}@di.uoa.gr.
}

\maketitle

\begin{abstract}
Semantic communication systems aim to transmit task-relevant information between devices capable of artificial intelligence, but their performance can degrade when heterogeneous transmitter--receiver models produce misaligned latent representations. Existing semantic alignment methods typically rely on additional digital processing at the transmitter or receiver, increasing overall device complexity. In this work, we introduce the first over-the-air semantic alignment framework based on stacked intelligent metasurfaces (SIM), which enables latent-space alignment directly in the wave domain, reducing substantially the computational burden at the device level. We model SIMs as trainable linear operators capable of emulating both supervised linear aligners and zero-shot Parseval-frame-based equalizers. To realize these operators physically, we develop a gradient-based optimization procedure that tailors the metasurface transfer function to a desired semantic mapping. Experiments with heterogeneous vision transformer (ViT) encoders show that SIMs can accurately reproduce both supervised and zero-shot semantic equalizers, achieving up to $90\%$ task accuracy in regimes with high signal-to-noise ratio (SNR), while maintaining strong robustness even at low SNR values.
 \smallskip
\end{abstract}


\section{Introduction} 
\label{sec:introduction}

The rapid proliferation of connected devices, coupled with the rise of latency- and data-intensive applications, is exposing fundamental limitations in traditional bit-centric communication architectures. Although systems grounded in Shannon's separation theorem remain highly effective for reliable bit transmission, they struggle to meet the stringent latency, bandwidth, and energy constraints imposed by autonomous systems, industrial automation, and large-scale Internet of Things~\cite{alwis2021devices}. These limitations have motivated increasing interest in semantic communications (SC) which prioritize the transmission of task-relevant information rather than exact bitwise representations \cite{strinati20216g, gunduz2022beyond, strinati2024goal}. By extracting and transmitting compact, semantic features tailored to the downstream task, SC can substantially reduce communication overhead and enhance energy efficiency. To this aim, deep neural networks (DNNs) are often used to extract low-dimensional, task-relevant latent features that replace conventional symbol streams, enabling communication directly at the level of meaning or task utility.

Two representative SC examples are Edge Inference (EI) and Deep Joint Source–Channel Coding (DJSCC). In EI, resource-constrained devices collaborate with nearby edge servers to execute DNNs through split or distributed inference~\cite{deng2020edge}. In DJSCC, neural encoders and decoders are trained end-to-end to jointly learn semantic compression and channel codes \cite{gunduz2024joint, bourtsoulatze2019deep, xu2023deep}. However, while DJSCC has demonstrated impressive performance, its underlying formulations generally assume that the transmitter (TX) and receiver (RX) operate within a shared latent space. This assumption is often violated in practice, as real-world devices tend to be heterogeneous employing independently trained models, distinct architectures, or protected designs constrained by privacy or intellectual property considerations. Such constraints prevent joint training or direct model exchange, and give rise to latent-space misalignment, a form of \textit{semantic noise} that can significantly degrade performance even when the physical channel itself is ideal \cite{luo2022semantic, sana2023semantic}. In this context, \emph{semantic alignment} (a.k.a. \emph{semantic channel equalization}) refers to techniques that align the latent representations of heterogeneous or independently trained models, enabling them to communicate consistently without requiring model sharing or joint retraining \cite{sana2023semantic}.

\textbf{Related Works}. Recent work on cross-model alignment has explored two main directions. \textit{Supervised} methods learn explicit linear or structured mappings between latent spaces \cite{merullo2022linearly, moayeri2023text, maiorca2023latent, lahner2024direct}, often leveraging tools such as orthogonal Procrustes analysis \cite{wang2008manifold}, communication-aware constrained optimization \cite{pandolfo2025latent,huttebraucker2025ris,di2025federated}, DNN–based mappings that align heterogeneous models in DJSCC settings \cite{pannacci2025semantic} or that perform alignment jointly with DJSCC~\cite{pandolfo2025latent}. These methods typically require exchanging \textit{Semantic Pilots} (SPs), pairs of aligned latent representations or task-specific exemplars exchanged between devices, to estimate the transformation between latent spaces and enable reliable cross-model communication. In contrast, \textit{unsupervised zero-shot} methods avoid pilot exchange entirely by constructing isometry-invariant latent representations \cite{moschella2022relative}. These techniques rely on \textit{Anchors}, compact sets of reference samples that each model processes independently. Even though different models generate different latent embeddings for the same anchor set, the geometric relations among those embeddings remain consistent up to an isometry. By expressing every latent vector relative to this anchor-based coordinate system, each device obtains a representation that is invariant to rotations, reflections, or permutations of its internal latent space \cite{moschella2022relative}. This enables reliable inter-model communication without parameter sharing or paired latent exchanges, and recent work extends anchor-based invariants to dynamic, multi-agent settings \cite{fiorellino2025frame}.

In both supervised and zero-shot approaches, semantic equalization requires deploying pre-aligners and/or post-aligners at the TX and RX, thereby increasing system complexity and computational burden. To alleviate this overhead, recent work has explored performing portions of the computation directly in the physical layer. In particular, stacked intelligent metasurfaces (SIM) have emerged as a promising hardware platform capable of implementing high-dimensional linear transformations directly over the air (OTA), thus offloading processing from resource-limited edge devices and reducing end-to-end latency \cite{AXN23_SIM}. By leveraging wave-domain computation, SIMs enable physical-layer operations such as beamforming, analog modulation, and linear inference to be executed without digital processing, making them an attractive candidate for SC architectures \cite{Stylianopoulos_GO, DeepOAC, GJZ24_SIM_TOC}. However, to the best of our knowledge, no prior work has explored the use of SIMs for over-the-air semantic alignment.

\textbf{Contributions}. In this paper, we propose the use of SIMs to perform semantic alignment via wave-domain computation. Specifically, our contributions are fourfold: \textit{i}) we provide the first demonstration that SIMs can perform semantic alignment fully OTA, without the need for dedicated digital processing at the devices; \textit{ii}) we show that SIMs can emulate both supervised linear semantic equalizers and zero-shot Parseval-frame-based operators, thus supporting interoperability between heterogeneous TX and RX models; \textit{iii}) we develop a gradient-based electromagnetic (EM) optimization framework that tunes the SIM response to accurately approximate a target semantic transformation in the complex domain; and \textit{iv}) we deliver the first systematic analysis of SIM design parameters (layer depth, metasurface resolution, and inter-layer spacing) showcasing how they impact downstream semantic-task accuracy, providing practical guidelines for SIM-enabled SC. Numerical experiments corroborate our findings, illustrating the practical advantages of using SIMs for semantic alignment in artificial intelligence (AI)-native communications.\footnote{The source code for the numerical evaluation is available at: \href{https://github.com/SPAICOM/semantic-alignment-via-sim.git}{https://github.com/SPAICOM/semantic-alignment-via-sim.git}}

\section{System Model}
\label{sec:system_model}

\begin{figure}[t]
    \centering
    \includegraphics[width=0.98\columnwidth, trim=190bp 60bp 190bp 50bp, clip]{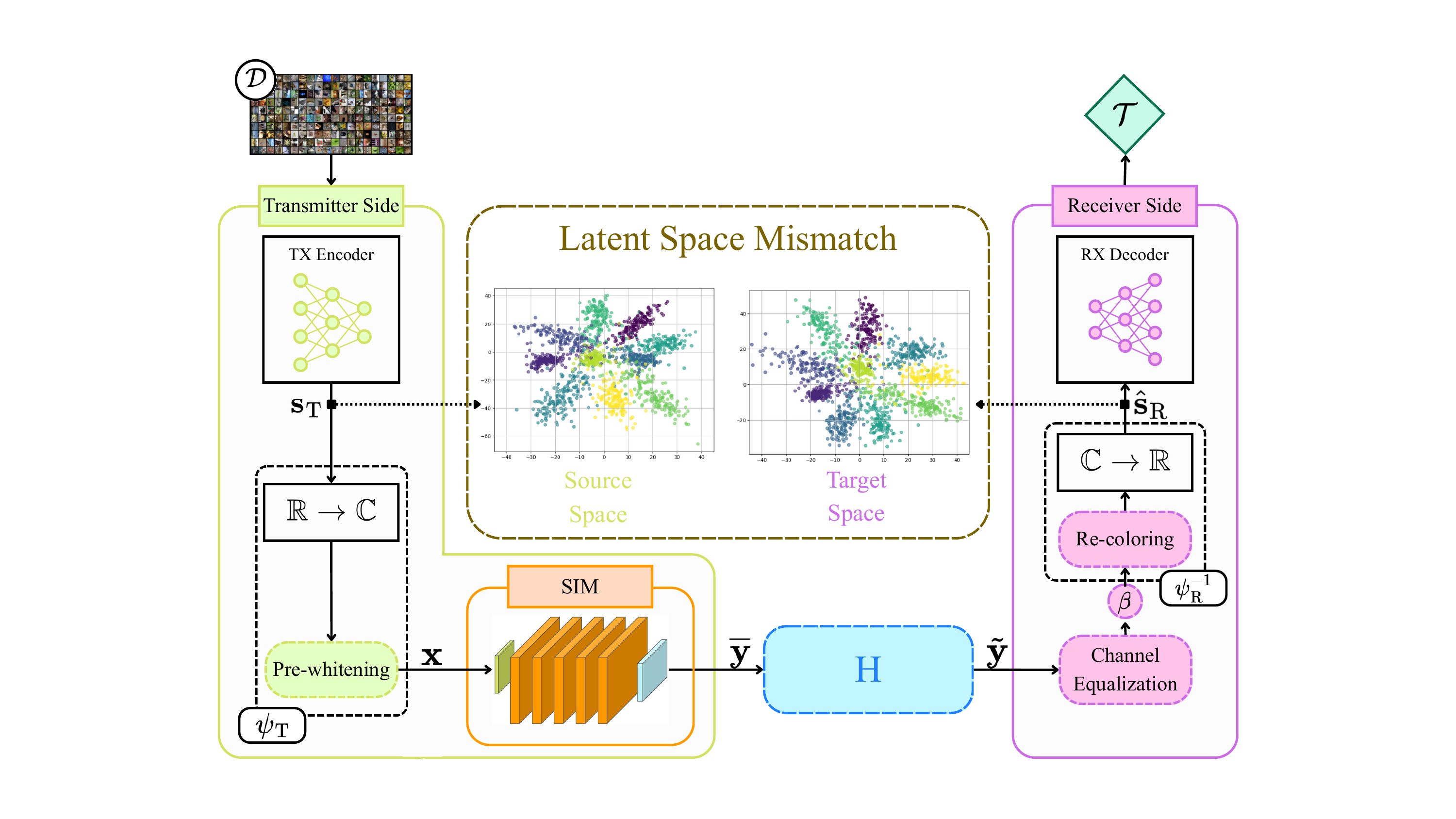}
    \vspace{-0.1cm}
    
    \caption{The proposed SC model: The SIM module performs OTA semantic equalization on the complex, compressed, and pre-whitened latent representation of TX before the transmission through a MIMO channel $\mathbf{H}$ with noise $\mathbf{v}$. At the RX side, channel equalization is first applied followed by decoding, in which the received signal is re-colored and decompressed to recover the message in the original RX latent space representation.}
    \label{fig:system_model}
    \vspace{-0.5cm}
\end{figure}

As illustrated in Fig.~\ref{fig:system_model}, we consider a multiple-input multiple-output (MIMO) SC framework in which a TX collaborates with an RX to perform a downstream task $\task$ (e.g., classification). These two agents employ \textit{pre-trained}, \textit{heterogeneous} DNNs for semantic encoding and decoding, enabling a possible exchange of latent representations. Specifically, let $\Tlatr \in \mathbb{R}^{\Tdim}$ denote the semantic feature vector extracted at the TX from a data sample $\delta \in \mathcal{D}$. The set of all such vectors defines the TX latent semantic space. Similarly, let $\Rlatr \in \mathbb{R}^{\Rdim}$ denote the semantic feature vector expected by the RX for the same data sample $\delta$ in order to correctly interpret the transmitted message and successfully perform $\task$. The collection of all $\Rlatr$ defines the RX latent semantic space. Since the TX and RX agents rely on heterogeneous DNN architectures, their latent spaces generally differ in both structure and dimensionality. As a result, the direct exchange of latent representations becomes susceptible to \textit{semantic noise}, leading to degraded performance in $\task$. Semantic alignment between heterogeneous latent spaces thus becomes necessary to enable mutual understanding between the two agents.

In our proposal, a SIM is integrated at the TX side, enabling the TX to transmit and align its latent representation, see Fig. \ref{fig:system_model}. To enable direct analog transmission of the latent features through the SIM, the real-valued latent vector must first be mapped into the complex domain. Assuming, without loss of generality, that the latent dimension $\Tdim$ is even, this mapping pairs the first half of the semantic features in $\Tlatr \in \mathbb{R}^{\Tdim}$ with the second half to form complex symbols. To further facilitate the SIM operation, a pre-whitening step is applied. The overall transformation, combining complex mapping and pre-whitening, is denoted by $\psi: \mathbb{R}^{2k} \!\to\! \mathbb{C}^{k}$. Since TX and RX rely on heterogeneous architectures, distinct mappings $\psi_\mathrm{T}$ and $\psi_\mathrm{R}$ are defined for each agent. These are applied to their respective latent representations $\Tlatr$ and $\Rlatr$, yielding the corresponding complex pre-whitened vectors $\x \in \mathbb{C}^{\Tdimc}$ and $\y \in \mathbb{C}^{\Rdimc}$, respectively, where $\Tdimc = \Tdim/2$ and $\Rdimc = \Rdim/2$.
The SIM can therefore be viewed as a parameterized, learnable mapping $\simmodule: \mathbb{C}^{\Tdimc} \to \mathbb{C}^{\Rdimc}$, optimized to perform semantic alignment during transmission between $\x$ and $\y$. We model this transformation as follows:
\begin{equation}
\overline{\y} = \simmodule(\x). \label{eq:sim_map}
\end{equation}
The communication between the TX and RX takes place through a multiple-input multiple-output (MIMO) wireless channel, modeled as a flat Rayleigh fading one represented by the matrix $\mathbf{H} \in \mathbb{C}^{\Nr \times \Nt}$, where $\Nt$ and $\Nr$ denote the number of TX and RX antennas, respectively. Each entry\footnote{We use notation $[\mathbf{z}]_i$ and $[\mathbf{Z}]_{i,j}$ to denote respectively the $i$-th element of a vector $\mathbf{z}$ and the $(i,j)$-th elements of a matrix $\mathbf{Z}$.} $[\mathbf{H}]_{i,j}$ is modeled as a zero-mean complex Gaussian random variable, accounting for the fading effect between the $i$-th TX and $j$-th RX antenna. The received signal is further corrupted by additive white Gaussian noise (AWGN) $\mathbf{v} \in \mathbb{C}^{\Nr}$, distributed as $\mathcal{CN}(\mathbf{0}, \sigma_v^2 \mathbf{I}_{\Nr})$. Then, assuming perfect channel state information (CSI) at the RX, channel equalization is performed using the minimum mean squared error (MMSE) equalizer $\equalizer = (\H^{H}\H + \tfrac{1}{\snr}\eye{\Rdimc})^{-1}\H^{H}$, where $\snr$ denotes the signal-to-noise ratio at the RX side. The equalized signal is then processed by the inverse mapping \(\psi_{\mathrm{R}}^{-1}: \mathbb{C}^{k} \!\to\! \mathbb{R}^{2k}\), which first re-colors the data and then converts it into real-valued form, thereby reconstructing the signal within the RX’s latent space. The overall semantic communication framework can therefore be summarized as:
\begin{equation}
    \hat{\mathbf{y}} = \equalizer\big(\mathbf{H}\, \simmodule(\mathbf{x}) + \mathbf{v}\big), \label{eq:com_summ}
\end{equation}
where the SIM module $\simmodule(\cdot)$ is optimized to perform OTA semantic alignment between TX/RX latent spaces. 

\subsection{SIM Model}
Consider an $L$-layer SIM collocated with the TX, as shown in Fig.\ref{fig:system_model}. The TX data are embedded in the wave domain through analog modulation, by configuring the input SIM layer as $\mathbf{\Upsilon}_0 = \amplconst_0 {\rm diag}(\x) \in \mathbb{C}^{\Tdimc \times \Tdimc}$, whose backplate is illuminated by a directive beacon signal as in \cite{LML23}, and $\amplconst_0$ is an optional amplification constant.
Each layer $l=1,\dots,L$ is assumed to have $M_l$ elements, with $M_0=\Tdimc$ and $M_L=\Nt$.
The element-to-element propagation between consecutive SIM layers is governed by geometric optics due to their dense placement~\cite{AXN23_SIM, Stylianopoulos_GO}.
Given elements $m$ and $m'$ ($1 \! \leq \! m \! \leq \! M_l$, $1 \! \leq \! m' \! \leq \! M_{l-1}$) with distance $d_{m,m'}$ and area $A_{\rm cell}$ from layers $l$  and $l\!-\!1$ ($2 \! \leq \! l \! \leq \! L$) of distance $s_{\rm layer}$, the propagation matrix $\mathbf{W}_l\! \in \! \mathbb{C}^{M_l \times M_{l-1}}$ can be expressed as:
\begin{align}
    [\mathbf{W}_l]_{m,m'} &= \frac{s_{\rm layer} A_{\rm cell}}{d_{m,m'}^2} 
    \Big( \frac{1}{2\pi d_{m,m'}} - \frac{\jmath}{\lambda} \Big) 
    e^{\jmath 2\pi d_{m,m'}},
\end{align}
where $\lambda$ is the carrier frequency and $\jmath\triangleq\sqrt{-1}$.
The responses of the unit elements of the $l$-th layer $\boldsymbol{v}_l \in \mathbb{C}^{M_l \times 1}$ are modeled as typical idealized phase shifters, i.e., $[\boldsymbol{v}_l]_m \triangleq \amplconst_l \exp(\jmath \xi_{l,m})$, where $\xi_{l,m}$ is the controllable phase shift and $\amplconst_l$ is a constant amplification term per element to compensate for attenuation in deep SIM structures.
By defining $\mathbf{\Upsilon}_l \triangleq {\rm diag}(\boldsymbol{v}_l)$, the overall SIM response can be mathematically expressed via the following matrix~\cite{AXN23_SIM}:
\begin{equation}
 \mathbf{G} =\prod_{l=1}^{L}  \mathbf{\Upsilon}_l \mathbf{W}_l \in\mathbb{C}^{M_L \times M_0},
\end{equation}
Therefore, the SIM-based mapping of~\eqref{eq:sim_map} is expressed as:
\begin{equation}
    \overline{\y} = \simmodule(\x) = \mathbf{G}\mathbf{\Upsilon}_0 \in \mathbb{C}^{\Nt \times \Tdimc}, \label{eq:sim_map-explicit}
\end{equation}
with $\boldsymbol{\xi} = \{\{\xi_{l,m}\}_{m=1}^{M_l}\}_{l=1}^{L}$ being the trainable parameters.

\section{SIM optimization for Semantic Alignment}
\label{sec:optimization}

In this section, we first formalize two semantic alignment strategies based on linear mappings, developed in: \textit{i}) a supervised and \textit{ii}) a zero-shot formulation. Subsequently, we introduce the SIM optimization procedure, which is designed to emulate these mappings over-the-air.

\subsection{Supervised Linear Semantic Alignment} 

A jointly optimized semantic equalizer, represented by the linear transformation 
$\AlignMatL \in \mathbb{C}^{\Rdimc \times \Tdimc}$, is designed to align the TX and RX latent spaces. 
The matrix $\AlignMatL$ is learned using a set of SPs, defined as a shared subset $\mathcal{S} \subset \mathcal{D}$ of reference data samples. Specifically, $\AlignMatL$ aligns the 
complex-compressed and pre-whitened latent representations of the samples in $\mathcal{S}$, given by 
$\mathbf{X} \in \mathbb{C}^{\Tdimc \times |\mathcal{S}|}$ at the TX and 
$\mathbf{Y} \in \mathbb{C}^{\Rdimc \times |\mathcal{S}|}$ at the RX. The alignment objective is formulated as a regularized least-squares problem  where the mean-squared error (MSE) between the two latent representations serves as a straightforward, yet effective measure of semantic mismatch, referred to as the \textit{Semantic Loss} $\SemL$:
\begin{equation}
    \min_{\AlignMatL}
    \underbrace{\left\lVert \mathbf{Y} - \AlignMatL \mathbf{X} \right\rVert_F^2}_{\text{Semantic Loss } \SemL}
    + \reg \left\lVert \AlignMatL \right\rVert_F^2,
    \label{eq:linear_prob}
\end{equation}
where $\reg \in \mathbb{R}_+$ is a non-negative regularization parameter introduced to improve numerical stability and control the magnitude of the alignment weights. Problem \eqref{eq:linear_prob} admits the following regularized least-square solution:
\begin{align}
\AlignMatL &= \Rlatc\Tlatc^H\left(\Tlatc\Tlatc^H+\reg\eye{\Tdimc}\right)^{-1}. \label{eq:linear_sol}
\end{align}
Clearly, the inclusion of the regularization term $\reg \eye{\Tdimc}$ guarantees numerical stability of the solution in \eqref{eq:linear_sol}. The resulting semantic equalizer provides the optimal linear alignment between the TX and RX latent spaces, learned in a supervised manner using SPs. While the SPs must be exchanged during training, the pre-whitened nature of both $\mathbf{X}$ and $\mathbf{Y}$ ensures that no original latent information is exposed, thereby preserving privacy. Once optimized, the equalizer enables effective semantic alignment between the two agents.

\subsection{Zero-Shot Parseval Frame Equalizers}

An unsupervised linear semantic equalizer avoiding SPs transmission can be implemented via \textit{Parseval Frame Equalizers} (PFEs)~\cite{fiorellino2025frame}. Let the TX and RX each be equipped with a pre-agreed, ordered set $\mathcal{A} = \{\alpha_1, \dots, \alpha_{|\mathcal{A}|}\}$, which contains the indices of the selected shared anchor data samples (e.g., images). Then, the private PFEs at TX and RX are:
\begin{equation}
    \mathbf{F}_{\mathrm{T}} = \mathbf{X}_{\mathcal{A}}\!\left(\mathbf{X}_{\mathcal{A}}^{H}\mathbf{X}_{\mathcal{A}}\right)^{-1/2}, 
    \quad 
    \mathbf{F}_{\mathrm{R}} = \mathbf{Y}_{\mathcal{A}}\!\left(\mathbf{Y}_{\mathcal{A}}^{H}\mathbf{Y}_{\mathcal{A}}\right)^{-1/2},
    \label{eq:pfe_def}
\end{equation}
where $\mathbf{X}_{\mathcal{A}} \in \mathbb{C}^{|\mathcal{A}|\times \Tdimc}$ and $\mathbf{Y}_{\mathcal{A}} \in \mathbb{C}^{|\mathcal{A}|\times \Rdimc}$ denote the private latent representations of the anchors points indexed in $\mathcal{A}$ at the TX and RX, respectively. The normalization in~\eqref{eq:pfe_def} ensures that both operators are well-conditioned, satisfying 
$\mathbf{F}_{\mathrm{T}}^{H}\mathbf{F}_{\mathrm{T}} = \eye{\Tdimc}$ 
and 
$\mathbf{F}_{\mathrm{R}}^{H}\mathbf{F}_{\mathrm{R}} = \eye{\Rdimc}$. When the cardinality of $\mathcal{A}$ is significantly larger than the respective latent dimensions (i.e., $|\mathcal{A}| \gg \Tdimc$ and $|\mathcal{A}| \gg \Rdimc$), the two operators form overcomplete Parseval frames. Conversely, when \(|\mathcal{A}| < \Tdimc\) and \(|\mathcal{A}| < \Rdimc\), the PFE operators \(\mathbf{F}_{\mathrm{T}}\) and \(\mathbf{F}_{\mathrm{R}}\) not only align the latent spaces but also compress the representations, while remaining perfectly conditioned within their respective spanned subspaces. Finally, the frame-based semantic equalizer $\AlignMatF \in \mathbb{C}^{\Rdimc \times \Tdimc}$ is obtained by composing the two PFE operators as follows:
\begin{equation}
    \AlignMatF = \mathbf{F}_{\mathrm{R}}^H\mathbf{F}_{\mathrm{T}}. \label{eq:pfe_sol}
\end{equation}
The quality of PFEs strongly depends on the suitability of the data samples indexed by $\mathcal{A}$. If the selected samples exhibit high linear dependence, the resulting operators may fail to adequately span the latent space. 
To address this issue, \cite{fiorellino2025frame} introduced the \textit{Proto Parseval Frame Equalizers} (PPFEs), which are based on the \textit{Prototypical Anchors} (PAs) selection strategy. 
The approach clusters the latent representations of the dataset $\mathcal{D}$ into $\nclusters$ groups, after which each prototypical anchor is obtained as the mean of $\nsamples$ randomly sampled latent vectors from its corresponding cluster, as detailed in Algorithm~\ref{alg:proto_alg}. In this construction, the number of clusters $\nclusters$ coincides with the number of anchors $|\mathcal{A}|$. In practice, clustering can be performed once on the latent space of a representative model and subsequently reused for other models. This semantic alignment strategy operates in a zero-shot fashion and exhibits strong numerical robustness. It typically requires only the pre-agreed, ordered sequence of anchors, thereby eliminating the need to transmit the SPs themselves. In the considered setting, however, the transmission of the RX's synthesis operator $\mathbf{F}_{\mathrm{R}}^{H}$ to the TX is also required, but incurs a communication cost proportional to $|\mathcal{A}|$, which is typically much smaller than the cost associated with transmitting $\mathbf{Y}$ in (\ref{eq:linear_sol}), whose size scales with $|\mathcal{S}|$.

\begin{algorithm}[t]
\caption{Prototypical Anchors}
\label{alg:proto_alg}
\begin{algorithmic}[1]
\STATE \textbf{Require:} $\mathcal{D}$, $\nclusters$, $\mathcal{A}$ \textbf{or} $\nsamples$, a neural encoder $E$, and a complex compression mapping $\psi$.
\STATE \textbf{Return:} Index set $\mathcal{A}$ and prototypical anchor matrix $\mathbf{P}$.
\IF{$\mathcal{A}$ is not provided}
    \STATE $\mathcal{X} \leftarrow E(\mathcal{D})$.
    \STATE $\{\cluster_1, \dots, \cluster_\nclusters\} \leftarrow$ apply a clustering algorithm with $\nclusters$ clusters to $\mathcal{X}$ such that $\bigcup_{i=1}^{\nclusters} \cluster_i = \mathcal{X}$.
    \STATE $\mathcal{A} = \{\mathcal{A}_1, \dots, \mathcal{A}_\nclusters\} \leftarrow$ for each cluster $\cluster_i$, randomly sample $\nsamples$ indices to form $\mathcal{A}_i$.
\ENDIF
\STATE Compute the prototypical anchors matrix as \\
$\mathbf{P} = \{\mathbf{p}_1, \dots, \mathbf{p}_\nclusters\}$, where each prototype is computed as:
$$
\mathbf{p}_i = \frac{1}{\nsamples} \sum\nolimits_{\alpha \in \mathcal{A}_i} \psi(\mathcal{X}_\alpha).
$$
\STATE \textbf{return} $\mathcal{A}$ and $\mathbf{P}$.
\end{algorithmic}
\end{algorithm}
\vskip -0.8cm

\subsection{SIM Optimization}

Fixing a semantic alignment matrix $\AlignMat$ as in (\ref{eq:linear_sol}) or (\ref{eq:pfe_sol}), we aim to optimize the EM response of the SIM such that it effectively emulates the linear transformation induced by $\AlignMat$. To this end, we define an emulation loss $\EmL$ measuring the discrepancy between the SIM’s EM response $\G$ and the target transformation $\AlignMat$, computed as the Frobenius norm between the following two matrices:
\begin{equation}
\EmL = \frob{\beta\G - \AlignMat}^2, \label{eq:emul_loss}
\end{equation}
where the scaling factor $\beta\in\mathbb{C}$ compensates for the overall amplitude attenuation in the SIM response. Following the formulation in~\cite{simDFT}, the optimization problem of minimizing the emulation loss of~\eqref{eq:emul_loss} can be expressed as:
\begin{subequations}\label{eq:sim_prob}
\begin{align}
    \min_{\{\xi_{l,m}\},\beta}\quad &\underbrace{\frob{\beta \G - \AlignMat}^2}_{\text{Emulation Loss } \EmL} \label{eq:sim_emulation_loss}\\
    \text{s. t.}\quad &\G = \Ups_L\W_{L}\dots\W_2\Ups_1\W_1, \label{eq:sim_const1}\\
    & \Ups_l = \text{diag}\left([e^{j\xi_{l,1}},e^{j\xi_{l,2}},\dots,e^{j\xi_{l,M_l}}]^T\right), \label{eq:sim_const2}\\
    & \{\xi_{l,m}\}_{m = 1,\dots,M_l}^{l = 1,\dots,L} \in [0,2\pi), \label{eq:sim_const3}
\end{align}
\end{subequations}
where~\eqref{eq:sim_const1}-\eqref{eq:sim_const3} characterize the SIM's EM response with respect to the controllable phase shifts. The optimization problem in~\eqref{eq:sim_prob} is inherently \textit{non-convex}, primarily due to the constant-modulus constraint on the metasurface phase shifts and the interdependence between layers~\cite{simDFT}. To address this, we employ a gradient-based iterative approach, where the phase parameters $\boldsymbol{\xi}_l = [\xi_{l,1}, \dots, \xi_{l,M_l}]$ are progressively adjusted to minimize the emulation loss $\EmL$. At each optimization step $t$, the phase updates follow the rule, as follows:
\begin{equation}
    \boldsymbol{\xi}_l^{(t+1)} \leftarrow \boldsymbol{\xi}_l^{(t)} - \eta \, \nabla_{\boldsymbol{\xi}_l}^{(t)} \EmL,
\end{equation}
where $\eta > 0$ denotes the learning rate. This iterative refinement allows the SIM to gradually tune its EM response to approximate the target transformation $\AlignMat$ with high fidelity. The scaling factor $\beta$ is also iteratively refined to preserve the proper magnitude of the SIM response. At each optimization step $t$, given the current estimate of the SIM transfer matrix, say $\G^{(t)}$, the optimal $\beta$ can be directly derived through the least-squares fitting procedure: 
\begin{equation}
    \beta = (\gvec^H\gvec)^{-1}\gvec^H \avec, \label{eq:beta_sol}
\end{equation}
where $\gvec=\text{vec}(\G^{(t)})$ and $\avec=\text{vec}(\A)$. The phase shift parameters are progressively refined until the emulation loss $\EmL$ stabilizes or the optimization reaches the prescribed iteration limit, balancing convergence accuracy and computational efficiency. Once the optimal pair $(\mathbf{G}, \beta)$ is obtained, the TX communicates the scalar factor $\beta$ to RX. Consequently, considering also complex-to-real (de-)whitening operations, the overall SC channel in~\eqref{eq:com_summ} can be cast as:
\begin{equation}
    \hat{\mathbf{s}}_\mathrm{R} = 
    \psi_\mathrm{R}^{-1}\!\left(
    \beta\, \equalizer\!\left(
    \mathbf{H}\mathbf{G}\psi_\mathrm{T}(\Tlatr) + \mathbf{v}
    \right)
    \right),
    \label{eq:comm_form_final}
\end{equation}
where the RX rescales the received signal by $\beta$, thereby achieving OTA semantic equalization through the SIM.

\section{Numerical Results}
\label{sec:numerical_results}

In this section, we evaluate the performance of the proposed semantic alignment methods through numerical experiments. The evaluation is conducted using the CIFAR-10 dataset, which contains $60\,000$ color images of size $32 \times 32$, evenly divided into 10 categories. About $70\%$ was used for training, $7.5\%$ was used for validation, while results are reported on the $12.5\%$ remaining test set. The downstream classification task is denoted by $\mathcal{T}$ and involves 10 target labels.
\begin{figure}[t]
    \centering    \includegraphics[width=\columnwidth, trim=5bp 5bp 5bp 5bp, clip]{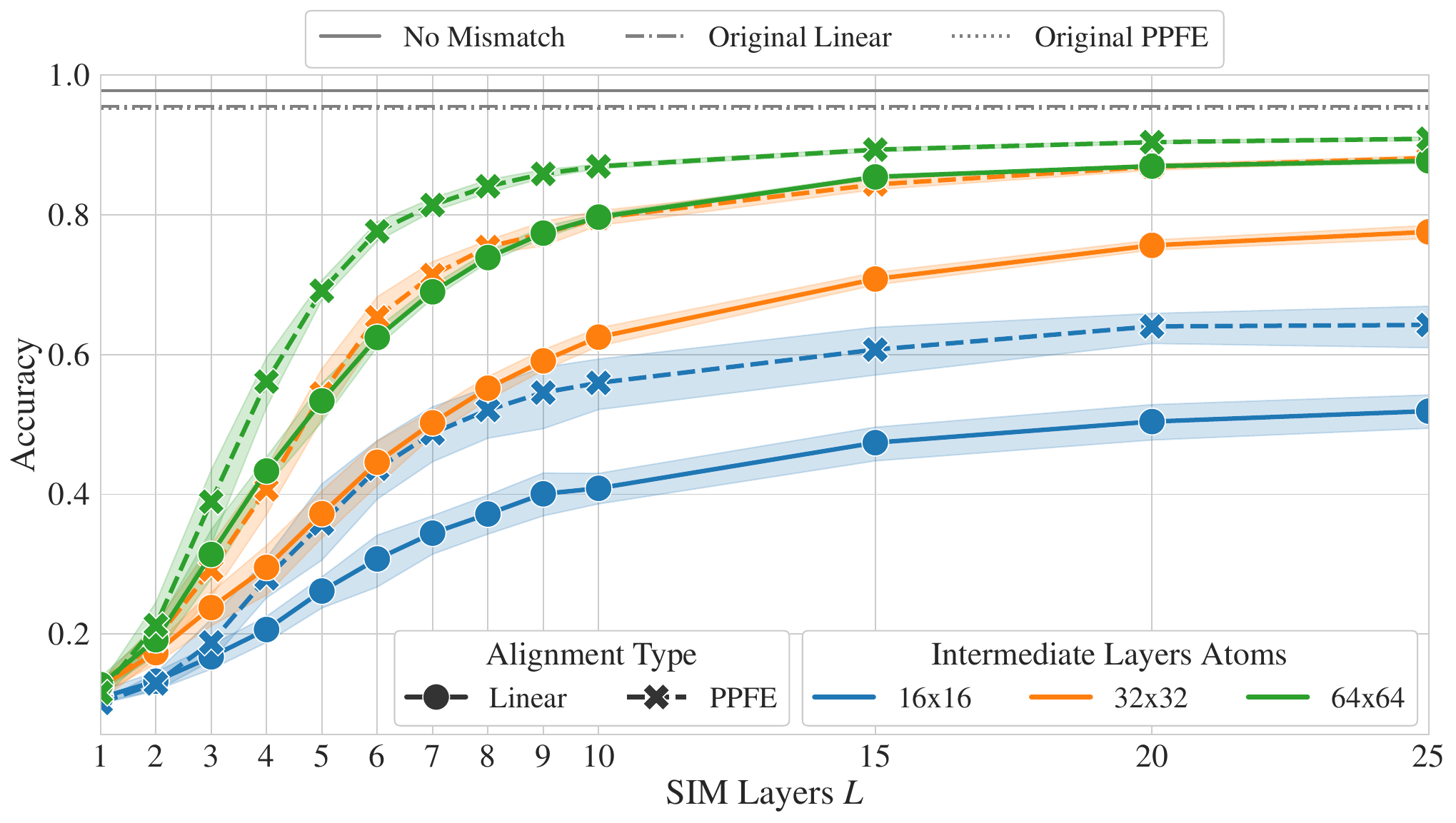}
    \caption{Accuracy versus $L$, considering infinite $\snrdb$.}
    \label{fig::accuracy_simlayers}
\vspace{-1em}
\end{figure}

For the encoding process, we utilize two pre-trained vision transformer (ViT) models available in the \textit{timm} library \cite{rw2019timm}. Specifically, the \texttt{vit\_small\_patch16\_224} model serves as the TX-side encoder with an embedding dimension of $\Tdim = 384$, while the \texttt{vit\_base\_patch16\_224} model is employed on the RX side, yielding encodings of dimension $\Rdim = 768$. We consider static square MIMO Rayleigh fading channels with unitary variance, assuming an equal number of transmit and receive antennas ($N_T = N_R$), both set to the dimension $\Rdimc$ corresponding to the complex compressed representation of the RX's latent space.
Each experiment is repeated using six different random seeds: $\{27, 42, 100, 123, 144, 200\}$. For every seed, a distinct channel realization $\mathbf{H}$ is generated under the same statistical assumptions. We fix $\reg = 0$, while the values of $\nclusters = 24$ and $\nsamples = 1000$ are selected through a grid-search procedure.
For the SIM, we set the width of each cell to $\lambda/2$, giving $A_{\rm cell} = \lambda^2/4$ with $\lambda=0.005$ m. Unless otherwise stated, we maintain an inter-element spacing of $s_{\rm layer} = 5\lambda$ and set $\amplconst_l = 0$ and each $\amplconst_l = 1$ to negate amplification. The learning rate is fixed to $\eta = 10^{-1}$, the number of iterations is set to 500, and the gradients are handled using the Adam optimizer. Each intermediate SIM layer is modeled as a rectangular array of $\sqrt{M_l} \times \sqrt{M_l}$ elements, where $M_l$ is kept constant across layers. The sizes of the first and last layers, $M_0$ and $M_L$, are determined by $\Tdimc$ and $\Rdimc$, respectively.

\begin{figure}[t]
    \centering    \includegraphics[width=\columnwidth, trim=5bp 5bp 5bp 5bp, clip]{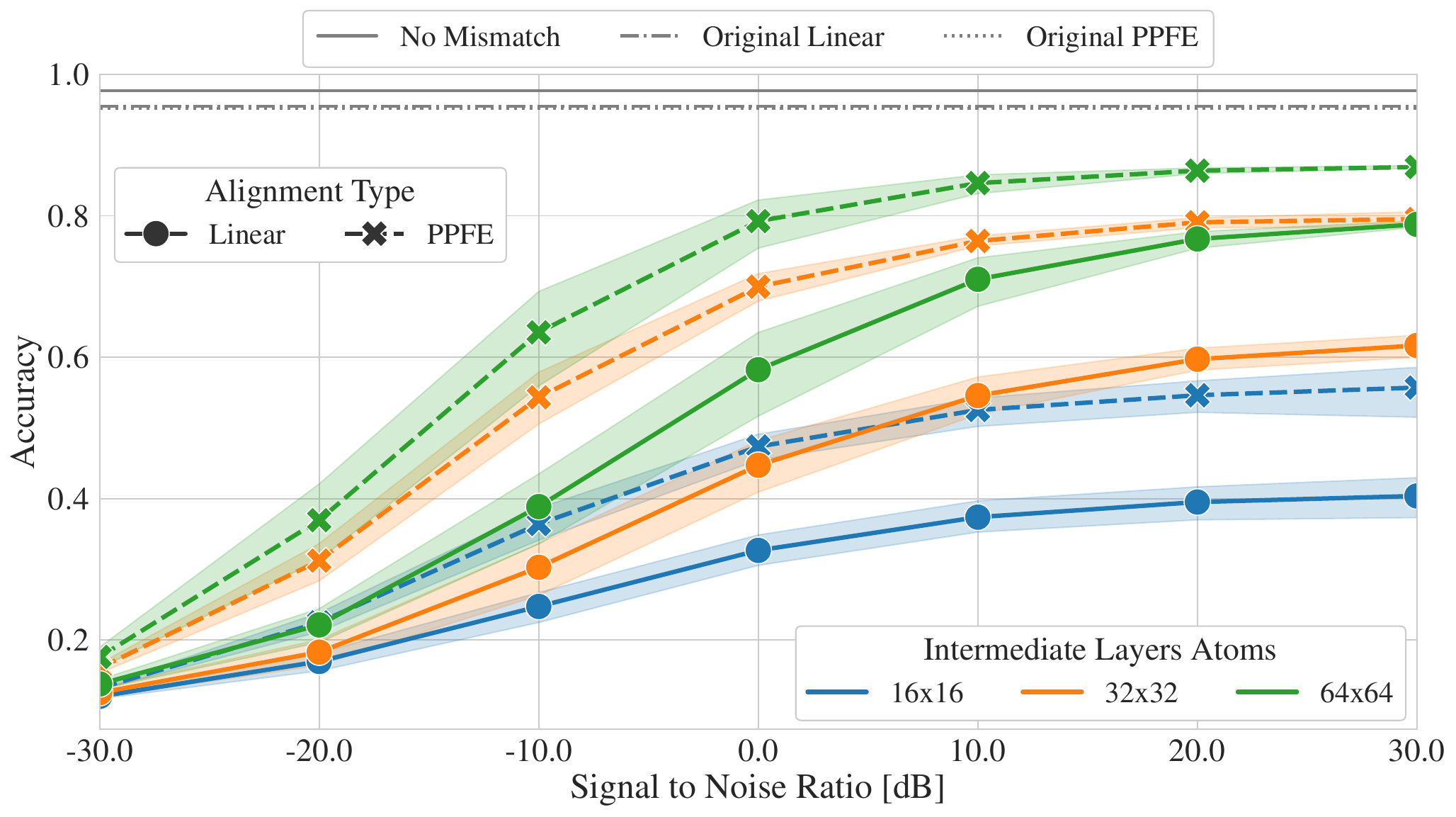}
    \caption{Accuracy versus $\snrdb$, considering $L=10$.}
    \label{fig::accuracy_snr}
\vspace{-1em}
\end{figure}

\begin{figure}[t]
    \centering    \includegraphics[width=\columnwidth, trim=5bp 5bp 5bp 5bp, clip]{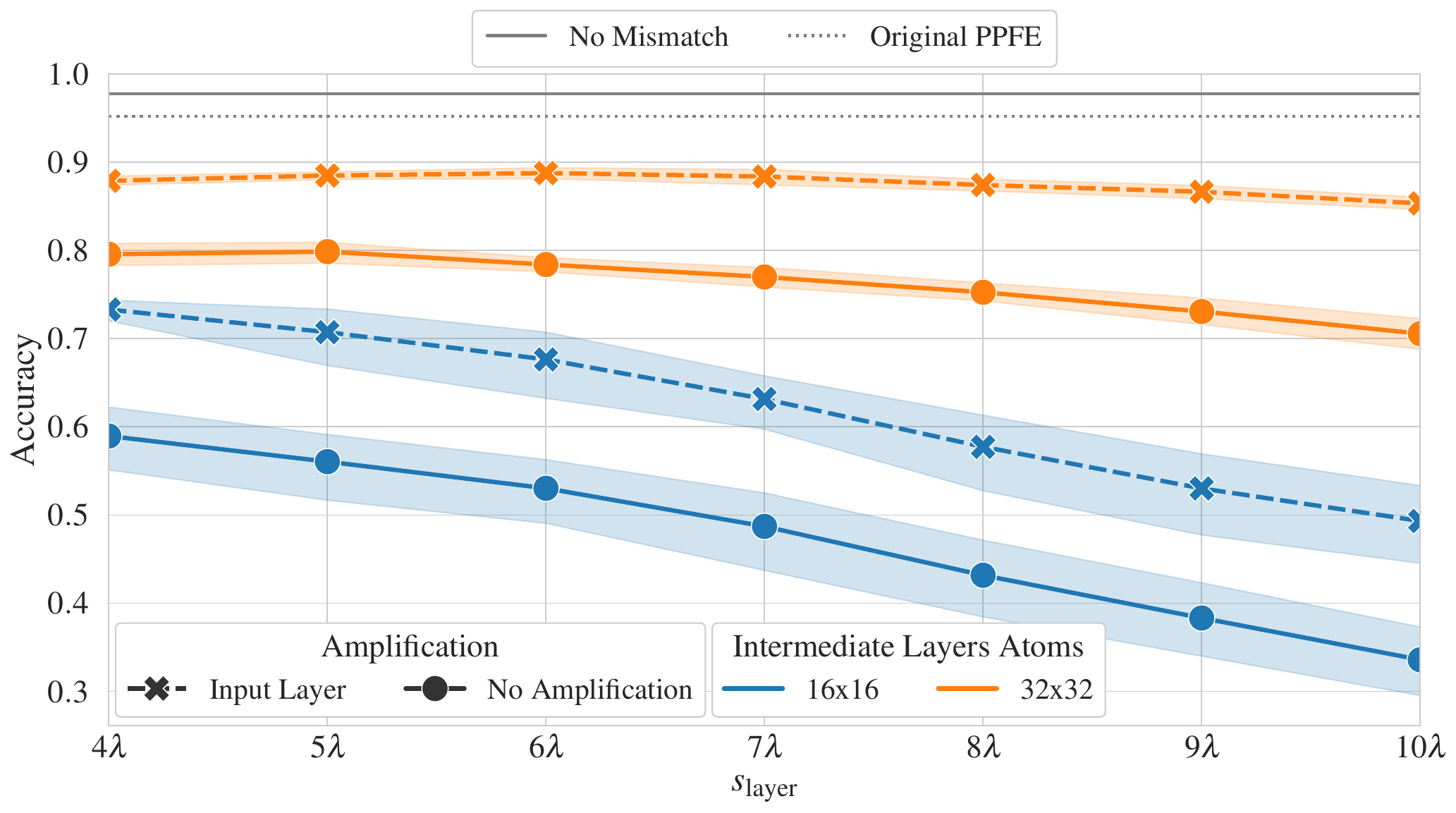}
    \caption{Accuracy versus $s_{\mathrm{layer}}$, considering only PPFE alignment.}
    \label{fig::accuracy_thickness_multiplier}
\vspace{-1em}
\end{figure}

In Figs.~\ref{fig::accuracy_simlayers}--~\ref{fig::accuracy_thickness_multiplier}, we report the performance of the downstream classification task under three reference conditions: \textit{i}) the case with no semantic misalignment (\textit{No Mismatch}); and \textit{ii}) the original, non–SIM-emulated semantic equalizers, namely the \textit{Original Linear} and \textit{Original PPFE} configurations. These original configurations serve as reference targets for the behaviors we aim to emulate, and they are displayed only when evaluating the SIM-based versions of the corresponding methods.

Figure~\ref{fig::accuracy_simlayers} reports the accuracy as a function of the number of SIM layers $L$ under an infinite $\snrdb$ regime, considering both linear and PPFE semantic equalizers across three different SIM configurations: $16 \times 16$, $32 \times 32$, and $64 \times 64$. These configurations correspond to the number of meta-atoms present in each intermediate layer of the SIM. The results demonstrate that increasing the size of the SIM configuration positively impacts the performance of the downstream classification task $\mathcal{T}$. Both the number of meta-atoms per intermediate layer and the number of SIM layers $L$ contribute significantly to this improvement. 
Specifically, the accuracy exhibits a pronounced improvement, increasing from slightly above \(60\%\) for the \(16 \times 16\) configuration to nearly \(90\%\) for \(32 \times 32\), and surpassing \(90\%\) for \(64 \times 64\). These levels are typically achieved around \(L = 20\) SIM layers.

Figure~\ref{fig::accuracy_snr} presents the classification accuracy across varying $\snrdb$ levels for the $16 \times 16$, $32 \times 32$, and $64 \times 64$ SIM configurations with $L = 10$ layers.
The results demonstrate that the SIM architecture maintains strong performance even at low SNRs: for the $64 \times 64$ configuration, it attains between 60–70\% downstream-task accuracy at $\snrdb = -10$~dB, and approximately 80\% at $\snrdb = 0$~dB when emulating the PPFE semantic equalizer. Moreover, the findings indicate that the SIM inherits the well-conditioned characteristics of the PPFE semantic equalizer, yielding greater robustness to noise relative to its linear-emulating counterpart. The PPFE-based SIM not only consistently outperforms the linear-emulating counterpart, but the $32 \times 32$ PPFE configuration can even surpass the $64 \times 64$ linear-emulating SIM in low-SNR regimes. Across all configurations, the performance improvement approaches saturation near \(\snrdb = 20\,\text{dB}\).

Figure~\ref{fig::accuracy_thickness_multiplier} reports the classification accuracy as the inter-layer spacing $s_{\mathrm{layer}}$ varies for SIM configurations of sizes $16\times16$ and $32\times32$, both employing $L = 10$ layers and evaluated with and without input-layer amplification (corresponding to $\amplconst_0 = 4/3$ and $\amplconst_0 = 1$, respectively). The results indicate that increasing $s_{\mathrm{layer}}$ generally degrades the performance of the classification task $\mathcal{T}$, although the impact becomes less pronounced for SIMs equipped with a larger number of intermediate layer atoms. Introducing a mild amplification at the input layer yields a substantial performance improvement, which, when combined with a reduced inter-layer spacing, enables smaller SIMs to emulate the performance of larger ones. Since $\beta$ effectively applies amplification at the output layer, additional amplification at that layer (i.e., setting $\amplconst_L > 1$) does not further influence the SIM’s task performance.

These findings show that the SIM can not only \emph{effectively emulate} both types of semantic equalizers for classification tasks, but also that it performs \emph{better} when emulating PPFE equalizers. In particular, the SIM consistently achieves higher downstream accuracy for $\mathcal{T}$ under PPFE emulation than under the Linear one, even though the original linear equalizer outperforms the original PPFE. Notably, PPFE equalizers impose an implicit compression determined by $\nclusters$ (i.e., $|\mathcal{A}|$), affecting both $\mathbf{F}_{\mathrm{T}}$ and $\mathbf{F}_{\mathrm{R}}$, a bottleneck absent in the linear case and yet the PPFE-emulating SIM still surpasses its linear-emulating counterpart. Moreover, SIMs equipped with fewer intermediate atoms but emulating PPFE equalizers perform comparably to, and in some cases rival, their linear-emulating counterparts that employ a larger number of intermediate atoms under the same system configurations.

\section{Conclusions}
\label{sec:conclusions}

This paper presented the first demonstration that SIM can perform semantic alignment entirely OTA, eliminating the need for dedicated digital processing at edge devices. We showed that SIMs can be optimized to emulate linear semantic aligners (both supervised and zero-shot PPFEs) thereby enabling heterogeneous TX and RX models to communicate reliably despite latent-space mismatches. Numerical results showcased that SIMs can accurately emulate the behavior of linear semantic aligners when equipped with sufficiently many layers and meta-atoms per layer, and when operating at high SNRs.
The study also provided systematic guidelines on the role of the SIM depth, layer size, and inter-layer spacing, revealing their substantial impact on semantic-task accuracy. Overall, these findings position SIMs as a powerful efficient hardware mechanism for scalable, energy-efficient, and AI-native semantic communications, opening the door to future research on jointly optimized physical-semantic programmable metasurface architectures.
\addtolength{\textheight}{-0.15in}

\bibliographystyle{IEEEtran}
\bibliography{refs}

\end{document}